\documentclass[a4paper,10pt]{article} 
\usepackage[utf8]{inputenc}
\usepackage{a4wide}
\usepackage{tgschola}
\usepackage{amssymb}
\usepackage{graphics,graphicx}
\usepackage[british]{babel}
\RequirePackage{doi}
\usepackage{hyperref}
\usepackage{url}
\usepackage{footmisc}
\usepackage{algorithm}
\usepackage{algpseudocode}
\usepackage{mathtools}
\usepackage{marginnote} 
\usepackage[table]{xcolor}
\usepackage{wrapfig} 
\usepackage[titletoc,page]{appendix}
\usepackage[affil-it]{authblk}
\usepackage{verbatim}
\usepackage{indentfirst}
\usepackage{paralist}

\DeclarePairedDelimiter{\ceil}{\lceil}{\rceil}
\floatname{algorithm}{Procedure}
\floatname{appendix}{Appendix}

\newcommand{\UPTO}{\textbf{upto} }
\newcommand{\AND}{\textbf{and} }
\newcommand{\OR}{\textbf{or} }
\newcommand{\TRUE}{\texttt{True} }
\newcommand{\FALSE}{\texttt{False} }

\title{Asynchronous Consensus Algorithm\footnote{referenced further as \emph{ACA}.}}

\author{Maxim Zakharov\thanks{\texttt{dp.maxime@gmail.com}}}
\affil{dpSearch Pty Ltd\footnote{partially done on contract with Offscale.io
    }
}
\date{\today}

\begin{document}
\maketitle

\abstract{
  This document describes a new consensus algorithm which is asynchronous and uses gossip based message dissemination between nodes.
  The current version of the algorithm does not cover the case of a node failure or significantly delayed response.
  This is the subject of further research of the algorithm.

  An outline of a new design for trust-less payment system is given in appendices.
}

\section{Introduction}

Consensus is a fundamental problem in distributed systems with a wide range of application. There are two types of consensus;
the first is \emph{Consensus on a value}, or, on the fact that an event has happened. The second is
\emph{Consensus on the linear order} of the events in the system. The later assumes a consensus on every value to be ordered and imposes
stricter requirement of the order of these events in the system.

The \emph{Lachesis protocol}~\cite{DBLP:journals/corr/abs-1810-02186,DBLP:journals/corr/abs-1810-10360} aims to provide consensus on linear order of events, however, implementation of it revealed several flaws\footnote{the concerns are given in appendices to this article}
that prevented it from reaching consensus at scale. Attempts to fix these issues led to a rethink of the approach which consequently produced the algorithm as described below.


\section{Acknowledgement}

It is a pleasure to thank Ann Marie Vande More whose help in improving the readability of this article is greatly appreciated.

\section {Basic notions}

\subsection{Node/peer} A node is an autonomous participant in the network of consensus (section~\ref{consensus-network}).
Usually, it is a standalone computer,
however, we consider a node to be any instance following this algorithm and having the following unique attributes
within the network:
\begin{itemize}
\item a pair of \emph{private/public keys};
\item its \emph{unique identifier}\footnote{for example, the first created public key of a node becomes its unique identifier};
\item a network address\footnote{usually it is an IP address and base port number\label{net-address-note}} at which this node operates following the algorithm.
\end{itemize}

\medskip

Other attributes of a node:
\begin{itemize}
\item current \emph{Lamport time} (section~\ref{lamport-time}) on the node;
\item current \emph{height} number, indicating the index of the last event created by the node;
\item current \emph{frame number} (section~\ref{frame-number});
\item last finalised frame number on the node;
\end{itemize}


In the description of the internal functionality of a node, the node is called a \emph{node}. In the description of node functions
as they are seen from another node, the node is called a \emph{peer}.
Otherwise the \emph{node} and the \emph{peer} are synonyms throughout this article.

\subsection{Peer list} The \emph{Peer List} is the list of all nodes that execute this algorithm and
combined into a single network
ensuring each peer is aware of the existence of all other peers of that network.
Unique attributes of each peer are known to every member of the network.

\subsection{Consensus network}\label{consensus-network} The \emph{Consensus network} is a set of nodes
that execute this algorithm and sharing common \emph{Peer List}.
In other words, they are aware of existance of each other and can communicate with
each other\footnote{the node connection graph may not be complete until it remains connected} following procedures of the algorithm.
The \emph{network} throughout this document means the \emph{Consensus network} unless otherwise specified.

\subsection{Lamport time}\label{lamport-time}
The \emph{Lamport time} is the virtual clock\footnote{see \cite{Lamport:1978:TCO:359545.359563} for the original idea description.}
of a node that follows specific rules:
\begin{enumerate}
\item There are two strategies to initialise it; the first is that all nodes start with the value set to 0, and the second
  is to initialise it with the value of 13th\footnote{or any other preselected} byte of the node's unique identifier.
  The former gives a higher number of events with the very same Lamport timestamp in initial rounds of the algorithm;
\item On creation of a new event, the node increases its Lamport time by 1 before assigning timestamp to the event;
  \item On synchronisation with other peer, the Lamport time is set to the maximum value of Lamport time on both peers.
\end{enumerate}

The \emph{Lamport timestamp} is a fixed value of the \emph{Lamport time}.

\subsection{Event}\label{event} The \emph{Event} is an atomic block of exchange between peers.
The event's \emph{creator} is a peer who created that event.
Each event has the following attributes:
\begin{itemize}
  \item its \emph{unique identifier};
\item creator's unique identifier;
\item creator's height index;
\item self-parent's unique identifier and \emph{hash value};
\item other-parent's unique identifier and \emph{hash value};
\item Lamport timestamp;
\item transaction payload (section~\ref{transaction-payload});
\item \emph{hash value} of all attributes above\footnote{these attributes are called \emph{hash domain attributes}.};
\item array of \emph{digital signatures} of the hash attribute above; one per each peer this event has been passed by;
\item \emph{frame number} (section~\ref{frame-number});
\item \emph{flag table} (section~\ref{flag-table}).
\end{itemize}
Please note, the \emph{frame number} and the \emph{flag table} attributes are not passed over the network (communicated between peers); each node of the network calculates values of these attributes independently.

\emph{ACA} does not require any particular hashing algorithm to be used for the hash values; the choice of such algorithm is on the implementation.

\medskip

Events of the network form a Merkle-like tree\footnote{see~\cite{merkle1987digital} for the original idea}: each event references two other events as its parent
events by hash values of these events; one parent event
is the most recently created event of the node, this event is called \emph{self-parent}, and the second is the last known event\footnote{known to the node at the time of event creation} created by any other peer, this event is called \emph{other-parent}.
\emph{Leaf events} (see section~\ref{leaf-event}) are the base for such reference recursion. This structure ensures integrity of all events passed through the network.

If an event $y$ is in a sub-tree of the Merkle-like tree described in the paragraph above
with an event $x$ as its root,
then the event $x$ \emph{sees} the event $y$ and the event $y$ is \emph{visible} for the event $x$.

When a peer creates an event it signs it using its private key.
When a peer receives an event following the \emph{Synchronisation Procedure} (Procedure~\ref{proc:node:syncreq}) it verifies
all digital signatures of the event using public keys of corresponding peers and then sign it with its private key.
This structure provides additional guarantee of the event integrity.


\subsubsection{Transaction payload}\label{transaction-payload}
The \emph{Transaction payload} is an array of user transactions (which could be empty) combined with
an array of internal transactions (which again could be empty).

Internal transactions are separated from user transactions because they are hidden from users and can be used for implementation of internal network operations such as to handle dynamic node participation and private/public keys change.
They could also disseminate information about events failed in the consensus or any other data needed to implement any additional node/peer functionality.

\subsubsection{Leaf event}\label{leaf-event} The \emph{Leaf event} is the first event of a peer. It has special status as follows:
\begin{itemize}
\item it is created once a peer is added to the network;
\item it has empty transaction payload;
\item its height index value is $0$ (index is equal to creator's height at the time of event creation);
\item its Lamport timestamp is set to the initial value of the node's Lamport time;
\item its parents' unique identifiers and hashes are set to zero;
\item its Flag table contains only that leaf event itself with value of the frame number equal
  to the current frame number at the time of peer addition\footnote{at the time of network initialisation the value of the current frame is $0$.}.
\end{itemize}

This special status allows all leaf events to be created on each participating node independently without the need to communicate those leaf events to other participants.
Every newly added peer devises leaf events of all other participants from the \emph{Peer List} of the network.

\subsection{Frame number and Frame}\label{frame-number}
The \emph{Frame number} is a characteristic of an event which initial value is calculated in the \emph{Event Insertion Procedure} (Procedure~\ref{proc:event:insert}).
\emph{Frame} is a set of all messages in the network having the same frame number; and also all events of the same frame come into final ordering altogether;
this means all events in a frame with a lower number precede all events in a frame with a higher number.

\subsection{Root majority}\label{sec:root-majority}
The \emph{Root majority} is a number
indicating threshold value for the number of visible roots from the current frame for an
event before that event becomes the root of a new frame.
This parameter regulates approximately the number of events of
the same creator in a single frame. This could be of any value strictly
between $1$ and $n$, with $n$ being the size of the \emph{Peer List}. The nearest integer to $\frac{(n + 3)}{3}$ could be a good initial value for
it. Additional research is required to see if this parameter could be a
per-node parameter (with modification of its value via internal
transactions).

\subsection{Root} \emph{Leaf events} are roots by default; any event
seeing \emph{Root majority} roots of the previous frame becomes a root of a new frame.

\subsection{Flag Table and Creator Flag Table}\label{flag-table}
The \emph{Flag Table} is a map used in the algorithm to track propagation of events between nodes.
It uses  unique identifiers (IDs) of visible roots as keys of the map and for each root it stores the frame number of that root. $\{event.ID: frame number\}$

The \emph{Creator flag table} is a map similar to the \emph{Flag table} in that both reference visible roots, but
\emph{Creator flag table} uses creator's unique identifier as key of the map instead of event unique identifier.
This map is used to detect the moment when a frame may be finalised on a particular node. $\{creator.ID: frame number\}$

\subsection{Visibilis} An event becomes a \emph{Visibilis} when the size of its creator flag table becomes equal to the number of peers in the network. Once an event becomes Visibilis on the current node, the current node executes the \emph{Frame Finalisation Procedure} (Procedure~\ref{proc:frame:finalise}).

\subsection{Gossip List} The \emph{Gossip list} stores Lamport timestamps and event unique identifiers for every peer
and indicates when each peer has communicated or seen most recently as well as its last known event.

The merging of two Gossip Lists is a simple procedure that creates a new Gossip List consisting of the
maximal value of Lamport timestamp and unique identifier of the most recent event for every peer out of these two lists.


\section {Node procedures}

The functionality of a node is defined by two main procedures \textbf{A}~(Procedure~\ref{proc:A}) and \textbf{B}~(Procedure~\ref{proc:B})
executing in parallel.

\smallskip

Procedure \textbf{A} periodically selects the next peer from its peer list and requests from it all events known by remote peer but unknown by the current node.

\pagebreak

\begin{algorithm}
  \caption{Procedure \textbf{A}}
  \label{proc:A}
  \scriptsize{
    \begin{algorithmic}[1]
      \Loop
      \State start heartbeat timer;
      \State select a peer, $P$, following \emph{Peer Selection Procedure} (Procedure~\ref{proc:peer:selection});
      \State execute \emph{Synchronisation Procedure} (Procedure~\ref{proc:node:syncreq}) with peer $P$;
      \State wait until heartbeat timeout.
      \EndLoop
    \end{algorithmic}
  }
\end{algorithm}

The heartbeat period in the Procedure \textbf{A} has no effect on reaching the consensus and can be set to an arbitrary value which throttles
the frequency of communications between peers and thus controls the load on the node; it can be omitted.

\smallskip

Procedure \textbf{B} is a listener which reply to synchronisation requests from other peers\footnote{
  any practical implementation better should allow several instances of procedure \textbf{B} to run in parallel, one per
  each remote peer connected; or create a pool of such procedures to balance load on the node.}.

\begin{algorithm}
  \caption{Procedure \textbf{B}}
  \label{proc:B}
  \scriptsize{
    \begin{algorithmic}[1]
      \Loop
      \State receive a \emph{Synchronisation Request} from a peer, $A$;
      \State execute \emph{Synchronisation Reply Procedure} (Procedure~\ref{proc:node:syncreply}) with peer $A$.
      \EndLoop
    \end{algorithmic}
  }
\end{algorithm}

\subsection{Next Peer Selection Procedure}

The selection of the next peer to communicate with in the Procedure \textbf{A} should aim to assist the construction of a well balanced
tree of events in the network and therefore
minimise the height of sub-trees spanning all nodes. It should take into account all known events and the topology of the tree they induce.

For simplicity one can take the following idealistic approach: let all peers be sorted by public key and there are $n$ peers in total,
with $n > 1$.
The $current$ is the index of the node in that sorted list.
If $r$ is the peer selection round number, whic is an internal static variable, then here is Procedure~\ref{proc:peer:selection} to select the index of the next peer to connect. 

\begin{algorithm}
  \caption{Next peer selection procedure}
  \label{proc:peer:selection}
  \scriptsize{
    \begin{algorithmic}[1]
      \Require initially $r \leftarrow ~ n \gg 1$
      \Comment{here $\gg$ is the operation of binary shift to the right}
      \State $next = (current + r) \pmod{n}$
      \Comment{here $\pmod{n}$ is modulo $n$ operation}
      \If{ $r > 1$ }
      \State $r \leftarrow ~ r \gg 1$
      \Else
      \State $r \leftarrow ~ n \gg 1$
      \EndIf
      \State \Return $next$
    \end{algorithmic}
  }
\end{algorithm}

Another approach would be to select the next peer randomly from the peer list excluding the current node
and the most recently contacted peer. This approach is closer to real byzantine-like behaviour of a node, which
does not follow prescribed
next peer selection procedure, rather, it leads to a less optimal peer synchronisation pattern.

\subsection{Synchronisation Procedure}\label{node:syncreq}

The \emph{Synchronisation Procedure} is an active part of event propagation between nodes.
As the first step, it sends the \emph{Synchronysation Request}
which is formed by the current \emph{Gossip List} and the current \emph{Lamport time} of the node.

Then, upon receiving the \emph{Synchronisation Reply} which consists of the remote \emph{Gossip List},
the remote \emph{Lamport time} value and the bundle of events\footnote{not yet known by the current node}.  All received events in the bundle
are processed one by one in the order of the bundle by executing the
\emph{Event Insertion Procedure} (Procedure~\ref{proc:event:insert}) for each event.

In the next step, the remote \emph{Gossip List} is merged into the current \emph{Gossip List} while the
\emph{Lamport time} value of the current node is set to either its current value or the remote \emph{Lamport time} value, whichever is greater.

Finally, any pending user transaction, internal transaction or a non-finalised event will initiate a new event
executing the \emph{Event Creation Procedure} (Procedure~\ref{proc:event:insert}).

\begin{algorithm}
  \caption{Synchronisation Procedure}
  \label{proc:node:syncreq}
  \scriptsize{
    \begin{algorithmic}[1]
      \Require $P$ -- remote peer selected following Procedure~\ref{proc:peer:selection}
      \Require $GossipList$ -- current \emph{Gossip List} of the node
      \Require $LamportTime$ -- current \emph{Lamport time} of the node
      \State send \emph{Synchronysation Request} to remote peer $P$;
      \State receive a \emph{Synchronisation Reply} $R$;
      \ForAll{$event \in R.bundle$}
      \State $EventInsertionProcedure(event)$
      \EndFor
      \State $GossipList \leftarrow merge(GossipList, R.GossipList)$
      \State $LamportTime \leftarrow max(LamportTime, R.LamportTime)$
      \If{$\exists$ pending user or internal transaction \OR $\exists$ non-finalised event}
      \State $EventCreationProcedure(P)$ \Comment{see section~\ref{event:creation}}
      \EndIf
    \end{algorithmic}
  }
\end{algorithm}

\subsection{Synchronisation Reply Procedure}\label{node:syncreply}

The \emph{Synchronisation Reply Procedure} is a passive part of event propagation between nodes and is executed upon receiving
a \emph{Synchronisation Request} from a remote peer.

As the first step, it extracts the remote \emph{Gossip List} from the \emph{Synchronisation Request} received, compares it with the
current \emph{Gossip List} of the node and creates a bundle of all known messages not known by the remote peer\footnote{these are events from each known peer whose \emph{Lamport timestamp} greater or equal to the value from corresponding coordinate in the remote \emph{Gossip List}.}.

Then it sends bundled events to the remote peer along with the current \emph{Gossip List} and the current \emph{Lamport timestamp}. All three form
the \emph{Synchronisation Reply}.

Finally, it sets the \emph{Lamport time} value of the current node to either its current value or the remote \emph{Lamport time} value, whichever is greater.

\begin{algorithm}
  \caption{Synchronisation Reply Procedure}
  \label{proc:node:syncreply}
  \scriptsize{
    \begin{algorithmic}[1]
      \Require \emph{Synchronisation Request}, $Req$, from a peer, $P$.
      \Require $GossipList$ -- current \emph{Gossip List} of the node.
      \Require $LamportTime$ -- current \emph{Lamport time} of the node.
      \Ensure \emph{Synchronisation Reply}, $Rpl$, is sent back to the peer $P$.
      \State $Rpl.bundle \leftarrow \varnothing$
      \ForAll{$event \in GossipList$ \AND $event \notin Req.GossipList$}
      \State $Rpl.bundle \leftarrow Rpl.bundle + event$
      \EndFor
      \State $Rpl.GossipList \leftarrow GossipList$
      \State $Rpl.LamportTime \leftarrow LamportTime$
      \State $LamportTime \leftarrow max(LamportTime, Req.LamportTime)$
      \State send $Rpl$ to the peer $P$
    \end{algorithmic}
  }
\end{algorithm}


\subsection{Event Insertion Procedure} The \emph{Event Insertion Procedure} is executed each time an event is inserted
into the local storage of a node, either in the \emph{Synchronisation Procedure} (Procedure~\ref{proc:node:syncreq}),
or in the \emph{Event Creation Procedure} (Procedure~\ref{proc:event:creation}).
It calculates the \emph{frame number} and the \emph{flag table} attributes of the event, which are not transmitted over the network between peers.
This procedure also checks the condition for frame finalisation on the current node
and finalises the frames with that condition met.

\medskip

As the first step, this procedure calculates the frame number for the inserting event and checks the condition if the event becomes root
using the following rules:
\begin{itemize}
\item If the frame number of the other-parent event is greater than the frame number of the self-parent event then the event becomes root and its
frame number is set to the frame number of other-parent event.

\item Or, if the frame number of the self-parent event is greater than the frame number of the other-parent event then the event is not the root and its
frame number is set to the frame number of self-parent event.

\item Otherwise,
  both parent events 
  have the same frame number; and in that case:
  \begin{itemize}
  \item the \emph{Root Flag Table}
  is calculated using the \emph{Strict Flag Table Merging Procedure}
  (see section~\ref{sec:flagtable-merging}) with the self-parent event frame number and
  the self-parent event flag table and the other-parent event flag table as parameters;
  \item and then the \emph{Creator Flag Table} is derived from
  the \emph{Root Flag Table} and the self-parent event's frame number using
  the \emph{Creator Table Derivation Procedure} (Procedure~\ref{proc:creator:table}).
  \item Now, if the size of the \emph{Creator Flag Table} is greater or equal to the value of \emph{Root Majority} (see section~\ref{sec:root-majority}),
  then the event becomes root and its frame number is set to one more than the self-parent event's frame number.
  \end{itemize}
\end{itemize}

Next, the \emph{Visibilis Flag Table} is calculated using
the \emph{Open Flag Table Merging Procedure} (Section~\ref{sec:flagtable-merging})
with the last finalised frame number plus one, the self-parent event flag table and the other-parent event flag table as parameters.
For the event detected as root in the previous step,
its unique identifier and the frame number are added into the \emph{Visibilis Flag Table}.
After that, the \emph{Visibilis Flag Table} becomes the flag table of the inserting event.
At this point the event could be stored in the local database.

\medskip

The final step is to check the condition for frame finalisation. To do so, the \emph{Creator Visibilis Flag Table} is devised from the
\emph{Visibilis Flag Table} and the number of first not yet finalised frame as parameters to the
\emph{Creator Table Derivation Procedure} (Procedure~\ref{proc:creator:table}). If the size of the \emph{Creator Visibilis Flag Table} is
equal to the size of the \emph{Peer List} the procedure finds the minimal value of the frame number in the \emph{Creator Visibilis Flag Table}
and finalises all frames up to this number by calling the \emph{Frame Finalisation Procedure} (Procedure~\ref{proc:frame:finalise})
consecutively for each frame to finalise. It is important that the frame with the number equal to the minimal found is not
finalised at this stage.

\begin{algorithm}
  \caption{Event insertion procedure}
  \label{proc:event:insert}
  \scriptsize{
    \begin{algorithmic}[1]
      \Require $event$ -- the event being inserted
      \Require $selfParent$, $otherParent$ -- parent events of the event being inserted
      \Require $lastFinalisedFrame$ -- the number of the last finalised frame on the node
      \Require $n$ -- the size of peer list, i.e. the nuumber of peers in the network
      \Ensure \emph{frame number} of the $event$ is calculated
      \Ensure \emph{flag table} of the $event$ is calculated
      \If{$selfParent.Frame = otherParent.Frame$}
      \State $rootFlagTable \leftarrow strictMergeFlagTables(selfParent.Frame, selfParent.FlagTable, otherParent.FlagTable)$
      \State $creatorRootFlagTable \leftarrow deriveCreatorTable(rootFlagTable)$
      \If{$length(creatorRootFlagTable) \ge rootMajority$}
      \State $root \leftarrow \TRUE$
      \State $frame \leftarrow selfParent.Frame + 1$
      \Else
      \State $root \leftarrow \FALSE$
      \State $frame \leftarrow selfParent.Frame$
      \EndIf
      \ElsIf{$selfParent.Frame > otherParent.Frame$}
      \State $root \leftarrow \FALSE$
      \State $frame \leftarrow selfParent.Frame$
      \Else
      \State $root \leftarrow \TRUE$
      \State $frame \leftarrow otherParent.Frame$
      \EndIf
      \State $event.Frame \leftarrow frame$
      \State $visibilisFlagTable \leftarrow openMergeFlagTables(lastFinalisedFrame+1, selfParent.FlagTable, otherParent.FlagTable)$
      \If{$root$}
      \State $visibilisFlagTable\{event.ID\} \leftarrow frame$
      \EndIf
      \State $event.FlagTable \leftarrow visibilisFlagTable$
      \State \Comment{... store $event$ into local database ...}
      \State $creatorVisibilisFlagTable \leftarrow deriveCreatorTable(visibilisFlagTable)$
      \If{$length(creatorVisibilisFlagTable) = n$}
      \State $frameToFinaliseUpto \leftarrow minFrameInFlagTable(creatorVisibilisFlagTable)$\label{line:minFrame}
      \For{$frame=lastFinalisedFrame+1$ \UPTO $frameToFinaliseUpto$}\label{line:upto}
      \State $frameFinalisationProcedure(frame)$
      \EndFor
      \EndIf
    \end{algorithmic}
  }
\end{algorithm}

Please note, it is important that the loop on the line~\ref{line:upto} is not executed for the value of $frameToFinaliseUpto$.

\medskip

Function $minFrameInFlagTable()$ on the line~\ref{line:minFrame} looks up for the smallest \emph{Frame number} in the \emph{CreatorFlag Table} given.


\subsection{Event Creation Procedure}\label{event:creation}
A new event may be created each time a node synchronises with another peer following \emph{procedure A} (Procedure~\ref{proc:A}).
The \emph{Event Creation Procedure} calculates attributes of the new events according to the following rules:
\begin{itemize}
\item The \emph{creator's unique identifier} is set to the unique identifier of the current node (that is, the one that creates the event);

\item The \emph{creator's height index} is set to the next index of the event created by the current peer\footnote{\emph{ACA}
  doesn't rely on uniqueness of this value for all events created by a peer, though this property is for convenience and provides additional
check for integrity of events from the same peer.};

\item The \emph{self-parent's hash} and \emph{unique identifier} are set to the hash and unique identifier of the last event
created by the current node\footnote{the new event is considered created when this procedure finishes.};

\item The \emph{other-parent's hash} and \emph{unique identifier} are set to the hash and unique identifier of the last known event of the peer just communicated following \emph{procedure A} (Procedure~\ref{proc:A});

\item The \emph{Lamport timestamp} of the new event is set to the next value of the node's Lamport time\footnote{node's Lamport time is increased by 1 before assigning to the event};

\item The \emph{transaction payload} is created from pending internal and external\footnote{external transactions -- those received from customers} transactions;

\item The \emph{hash value} is calculated as hash (control sum) of the values of all attributes above;

\item The \emph{signature} of the hash value  above (or all attributes above) is created using node's private key and put it into signatures array;
\end{itemize}
After all attributes above are filled, this procedure executes the \emph{Event insertion procedure} (Procedure~\ref{proc:event:insert})
for the created event, which calculates the value of the \emph{frame number} and the \emph{Flag table} for created event.

\begin{algorithm}
  \caption{Event creation procedure}
  \label{proc:event:creation}
  \scriptsize{
    \begin{algorithmic}[1]
      \Require $peer$ -- a remote peer as other-parent
      \Require $node$ -- current node
      \Ensure Attributes of the new event are filled.
      \State $event \leftarrow \varnothing$ \Comment{a new object $event$ is created}
      \State $event.signatures \leftarrow \varnothing$
      \State $event.creatorID \leftarrow node.ID$
      \State $lastSelfEvent \leftarrow getLastEvent(node.ID)$\label{line:getLastEvent1}
      \State $lastOtherEvent \leftarrow getLastEvent(peer.ID)$\label{line:getLastEvent2}
      \State $node.height \leftarrow node.height + 1$
      \State $event.height = node.height$
      \State $event.selfParent.ID \leftarrow lastSelfEvent.ID$
      \State $event.selfParent.hash \leftarrow lastSelfEvent.hash$
      \State $event.otherParent.ID \leftarrow lastOtherEvent.ID$
      \State $event.otherParent.hash \leftarrow lastOtherEvent.hash$
      \State $node.lamportTime \leftarrow node.lamportTime + 1$
      \State $event.lamportTimestamp \leftarrow node.lamportTime$
      \State $event.payload \leftarrow createPayload()$\label{line:createPayload}
      \State $event.hash \leftarrow Hash(event)$\label{line:Hash}
      \State $event.signatures \leftarrow event.signatures + node.Sign(event)$\label{line:Sign}
      \State $EventInsertionProcedure(event)$
    \end{algorithmic}
  }
\end{algorithm}

Function $getLastEvent()$ on the lines \ref{line:getLastEvent1} and \ref{line:getLastEvent2} retrieves the last known event
for the peer with specified unique identifier.

Function $createPayload()$ on the line~\ref{line:createPayload} creates payload for the event out of pending internal and external
transactions.

Function $Hash()$ on the line~\ref{line:Hash} calculates hash (control sum) of the event over hash domain attributes.

Method $node.Sign()$ on the line~\ref{line:Sign} creates digital signature of the event hash value or of hash domain attributes and the
hash value.

\subsection{Flag Table Merging Procedures}\label{sec:flagtable-merging}
\begin{enumerate}
\item{\emph{Open procedure}\footnote{referred as $strictMergeFlagTable()$ in Procedure~\ref{proc:event:insert}.}:\\Open flag table merging procedure takes two flag tables and the frame number as parameters and forms a new flag table.
  This contains only the entries from source flag tables whose corresponding frame number is equal to or greater than the frame number specified.}
\item{\emph{Strict procedure}\footnote{referred as $openMergeFlagTable()$ in Procedure~\ref{proc:event:insert}.}:\\Strict flag table merging procedure takes two flag tables and the frame number as parameters and forms a new flag table.
  This contains only the entries from source flag tables whose corresponding frame number is equal to the frame number specified.}
\end{enumerate}

\subsection{Creator Table Derivation Procedure}
The \emph{Creator Table Derivation Procedure} is an auxiliary procedure used in the \emph{Event Insertion Procedure} (Procedure~\ref{proc:event:insert}). This procedure takes a flag table as an input and produces a map which stores the creator's unique identifiers of visible roots, and for each visible root it stores the minimal frame number.
\begin{algorithm}
  \caption{Creator table derivation procedure}
  \label{proc:creator:table}
  \scriptsize{
    \begin{algorithmic}[1]
      \Require $inputFlagTable$ -- flag table
      \Require $minFrameNumber$
      \State $resultCreatorTable \leftarrow \{\}$
      \ForAll{pair $\{eventID, frameNumber\} \in inputFlagTable$ \AND $frameNumber \ge minFrameNumber$ }
      \State $event \leftarrow getEvent(eventID)$\label{line:getEvent}
      \State $creator \leftarrow event.creatorID$
      \If{$\nexists resultCreatorTable\{creator\}$}
      \State $resultCreatorTable\{creator\} \leftarrow frameNumber$
      \ElsIf{$resultCreatorTable\{creator\} > frameNumber$}
      \State $resultCreatorTable\{creator\} \leftarrow frameNumber$
      \EndIf
      \EndFor
    \end{algorithmic}
  }
\end{algorithm}

The function $getEvent()$ on the line~\ref{line:getEvent} retrieves an event by its unique identifier.

\subsection{Frame Finalisation Procedure}\label{frame:finalise}

The \emph{Frame Finalisation Procedure} is called from the \emph{Event Insertion Procedure} when the condition for the frame finalisation is met
on the current node.

\medskip

Firstly, all events in the frame are sorted according the following rules:
\begin{enumerate}
  \setlength{\itemsep}{0pt}
  \setlength{\parskip}{0pt}
  \setlength{\parsep}{0pt}
\item the smaller Lamport timestamp has priority;
\item to break ties above, the smaller Lamport timestamp of self (grand-)$*$parents (recursively up to leaf events) has priority\footnote{For a large number of peers in the network, this rule will require significant amount of storage access operations and thus could be omitted or relaxed to self-parent's Lamport timestamp only due to performace reasons.};
\item to break ties above, the smaller hash value has priority;
  \item to break ties above in a rare case of hash collision, the smallest unique identifier has priority.
\end{enumerate}

Then each event in the frame is finalised in the order by executing \emph{Event Finalisation Procedure} (see Section~\ref{event:finalise}).

\begin{algorithm}
  \caption{Frame Finalisation Procedure}
  \label{proc:frame:finalise}
  \scriptsize{
    \begin{algorithmic}[1]
      \Require $Frame.events$ -- all events in the frame.
      \State $Sort(Frame.events)$\label{line:Sort}
      \ForAll{$event \in Frame.events$}
      \State $eventFinalisationProcedure(event)$
      \EndFor
     \end{algorithmic}
  }
\end{algorithm}

\subsection{Event Finalisation Procedure}\label{event:finalise}

The \emph{Event Finalisation Procedure} is called by the \emph{Frame Finalisation Procedure} for each event in the frame being
finalised\footnote{this procedure is called $eventFinalisationProcedure()$ in Procedure~\ref{proc:frame:finalise}.}.

For each event, the payload is processed in the following order:
\begin{inparaenum}
\item the external transactions are pushed to customers;
\item then internal transactions are processed;
\item and finally, the flag table is stripped off\footnote{to save storage space; though this step is optional.}.
\end{inparaenum}

Thus external transactions should not depend on the execution results of internal transactions from the same event, if they do, such external transactions must be
put into the next event of the creator.


\begin{appendices}

  \section*{Rust implementation}
  The \emph{ACA} has been implemented in the \emph{Rust} language. The implementation is available on Github: \url{https://github.com/Fantom-foundation/libconsensus-dag}\footnote{commit \texttt{1f9ec3570c70d51c060cfc5eba8d76f938890dcb}}


  This implementation of \emph{ACA}'s procedures relies on
  hashes and digital signatures as unique identifiers of events in the network,
however, none of hashing and digital signature algorithms provides guarantee of being collision-free.
Additional research is required to tailor modifications of the implementation
to make it collision proof.
Such research and application is at the discretion of the user.

  \section*{A note on Lachesis protocol and Swirlds algorithm}
  The Lachesis protocol~\cite{DBLP:journals/corr/abs-1810-02186,DBLP:journals/corr/abs-1810-10360} works with $n$ nodes and
  connects each new event with parent events from $k$ other nodes.
  Thus, a $k$-arny tree with $n$ leaf vertexes must be constructed before a node would be aware of all events of the same frame from
  all $n$ nodes. It is well-known that the minimal height for a tree with $n$ leaf vertexes is $\ceil{\log_k n}$.

  When a root becomes clotho it sees $\frac{2}{3}$ of roots of the previous frame,
  this means a tree of height at least $\ceil{log_k \frac{2}{3} n}$ should be constructed,
  but this value is less than or equal to the minimal height mentioned in the paragraph above.

  The atropos time selection procedure is executed straight after clotho status is confirmed (see Algorithm 1 in \cite{DBLP:journals/corr/abs-1810-10360}), thus, there is a possibility that atropos time will be selected before a node sees events of the same frame from all $n$ nodes.
  This means a non-zero probability that different nodes would select different atropos time (having different sets of events included into $\frac{2}{3}$ of $n$ peers seen be each node).

  \medskip

  The Swirlds hashgraph consensus algorithm~\cite{baird2016swirlds} is very similar to the Lachesis protocol,
  its \emph{divideRounds}, \emph{decideFame}, \emph{findOrder}
  procedures are executed after each reception of a sync, and in the same way as for the Lachesis protocol there is a non-zero probability that different nodes would select
  different events in the next round received and thus sorted them differently in the return of \emph{findOrder} procedure
  because the Swirlds algorithm requires $\frac{2}{3}$ of witness events\footnote{one witness event per network member
    in each round},
  and this is very same case of tree heights as for the Lachesis protocol.

\medskip

  Both the Lachesis protocol and the Swirlds consensus algorithm have the very same design flaw:
  they lack a mechanism of detecting if a particular node has received
  all events of a particular round/frame from all other nodes and thus would execute voting and produce
  the final event sorting prematurely. This problem will aggravate with the growth of network size, that is, the number of participating nodes.

  \section*{TxFlow: a new electronic payment system}

  In the seminal paper \emph{Bitcoin: A Peer-to-peer Electronic Cash System}~\cite{nakamoto2008bitcoin}, Satoshi Nakamoto
  outlined the design of a new payment system when the trusted third party has been replaced with the cryptographic proof and the public
  disclosure of all operations with the electronic cash in order to solve the coin double-spending problem.

  The events of the \emph{ACA} defined in this article form Merkle-like tree (see section~\ref{event}),
  which provides the same level of the cryptographic proof as in the Bitcoin design. Making the whole system public,
  one can build a distributed payment system similar to the bitcoin network but without a need to form
  blocks of transactions. Below is the outline of such system design.

  \subsubsection*{Accounts}
  A peer with public and private keys (and so with an unique peer ID) represents an acount.
  Only owner of the private
  key can authorise operations on a particular account. Note, in this schema a peer may not execute ACA procedures
  and do supply/receive transactions over its account via another peer willing to do so.

  \subsubsection*{Transactions}
  A transaction in such a system could be anything put into transactions fiield of the ACA event. It could be a simple
  instruction to move funds from the account to another or a group of them;
  it could be a program for a
  virtual machine to execute a smart contract. We leave the semantic of transactions to the implementation.

  Overall, the ACA guarantees the key property of the system: each node receives all transactions in
  the very same order. In other words, every account receives the same flow of all transactions in the system,
  and having the same starting values for all accounts each node will have the very same state of all accounts after
  each transaction processed.
  This property solves double spending problem in a way that, out of two transactions spending the same funds,
  one would be delivered first for every participant and it would be the same transaction for all participants.
  This means everyone will accept and reject the very same transaction out of two conflicting transactions.


\end{appendices}

\pagebreak

\bibliographystyle{habbrv-doi.bst}
\bibliography{asynchronous-consensus-algorithm}

\end{document}